\documentclass[preprint,review,12pt]{elsarticle}

\usepackage{amssymb}
\usepackage{amsmath}

\journal{Journal of Process Control}

\begin{document}

\begin{frontmatter}



\title{Fault Detection and Identification Using a Novel Process Decomposition Algorithm for Distributed Process Monitoring}

\author[label1]{Enrique Luna Villagómez}

\author[label1]{Vladimir Mahalec\corref{cor1}}
\ead{mahalec@mcmaster.ca}
\cortext[cor1]{Corresponding author}

\affiliation[label1]{organization={McMaster University, Department of Chemical Engineering}, 
            addressline={1280 Main St W}, 
            city={Hamilton},
            postcode={L8S 4L8}, 
            state={Ontario},
            country={Canada}}




\begin{abstract}
Recent progress in fault detection and identification increasingly relies on sophisticated techniques for fault detection, applied through either centralized or distributed approaches. Instead of increasing the sophistication of the fault detection method, this work introduces a novel algorithm for determining process blocks of interacting measurements and applies principal component analysis (PCA) at the block level to identify fault occurrences. Additionally, we define a novel contributions map that scales the magnitudes of disparate faults to facilitate the visual identification of abnormal values of measured variables and analysis of fault propagation. Bayesian aggregate fault index and block fault indices vs. time pinpoint origins of the fault. The proposed method yields fault detection rates on par with most sophisticated centralized or distributed methods on the Tennessee Eastman Plant benchmark. Since the decomposition algorithm relies on the process flowsheet and control loop structures, practicing control engineers can implement the proposed method in a straightforward manner.  
\end{abstract}





\end{frontmatter}



\section{Introduction}
\label{intro}

Equipment malfunctions or instrument faults significantly degrade the performance of industrial plants, leading to abnormal operating conditions that threaten both safety and profitability. As a result, monitoring systems are now standard in most industrial processes to prevent faults from escalating into failures \cite{Isermann1997}. Despite the advancements in safety protocols and monitoring technologies, recent studies indicate that incidents remain prevalent within the chemical industry \cite{Tamascelli2024}. This ongoing issue highlights the need for better process monitoring systems that can handle the complexities of modern large-scale operations while remaining simple for control engineers to implement and maintain. 

Process monitoring methods are typically classified into model-based, knowledge-based, and data-driven approaches \cite{Venkatasubramanian2003}. The limitations of model-based and knowledge-based frameworks in large-scale industrial systems, along with the increased availability of process data, have led to a surge in data-driven process monitoring research \cite{Ren2019,Sun2020,Bi2021,Lomov2021,Cacciarelli2022}.

Data-driven monitoring systems can be deployed in either a centralized or distributed configuration. Centralized systems consolidate process data into a single fault detection entity, simplifying data management and yielding good fault detection and identification (FDI) performance in small systems \cite{Ge2016}. However, in processes that involve complex interactions among numerous process variables, centralized systems can miss critical local information needed for accurate fault diagnosis, as faults are often caused by only a subset of these variables \cite{Yin2022}.

To address this limitation, distributed process monitoring systems are designed to decentralize monitoring tasks by focusing on specific process segments independently. This configuration improves the accuracy and responsiveness of the monitoring system by isolating faults in localized areas and reducing latency \cite{Jiang2019}. A critical step in implementing a distributed monitoring system is forming well-structured monitoring blocks that encompass variables with strong interactions to accurately represent localized aspects of the process \cite{He2020}. The construction of these monitoring blocks, referred to as process decomposition, can be accomplished using knowledge-based or data-driven methods. 

Knowledge-based approaches use domain expertise to incorporate process topology and significant variable relationships when forming monitoring blocks \cite{Qin2001}. Conversely, data-driven approaches rely on statistical features of process data, such as variable correlations or mutual information, to construct monitoring blocks \cite{Jiang2019}. Although data-driven distributed monitoring offers a promising approach to address the challenges of large-scale, complex industrial systems, many process decomposition techniques require specialized training or a deep understanding of the process to be implemented effectively and may not be easily accessible to the average control engineer.

Extensive research has been conducted on implementing distributed monitoring systems using various knowledge-based process decomposition approaches. For example, Grbovic et al. \cite{Grbovic2012} based process decomposition on sensor proximity to units and streams, which resulted in more monitoring blocks than process units. A similar approach was used by Zhu et al. \cite{Zhu2017}, who developed a distributed parallel PCA method using the process flow diagram (PFD) to define monitoring blocks for each process unit. Through process expertise, they merged the monitoring blocks with fewer sensors to ensure that the blocks have sufficient information. Chen et al. \cite{Chen2019} proposed a Distributed Canonical Correlation Analysis (DCCA) monitoring framework, enhancing fault detection by considering real-time inter-unit correlations. Despite using PFD information to guide the process decomposition, blocks with few measurements were merged without ensuring continuity or shared control dependencies. Subsequently, to address the dynamic and nonlinear nature of large-scale processes, Zou et al. \cite{Zou2024} implemented a Distributed Temporal-Spatial Enhanced Variational Autoencoder (DTSEVA) for plant-wide monitoring, enhancing fault detection by independently extracting spatial and temporal features. Their approach relied on expert knowledge for process decomposition, selectively merging low-information blocks while leaving others intact.

Meanwhile, Wu et al. \cite{Wu2022} proposed an alternative approach that uses digraph partitioning to define monitoring blocks. A digraph of causal relationships among process variables is constructed using process expertise, and a fast-unfolding algorithm \cite{BlondelVincentD2008}partitions the graph into blocks. Fault detection is then performed with temporal-spatial distributed canonical variate analysis (TS-DCVA), capturing temporal and spatial information for improved accuracy. Building on this concept, Lagare et al. \cite{Lagare2023} represented the PFD as a digraph, using units and streams as vertices rather than process variables, and created monitoring blocks centered on vertices where faults had previously occurred.

Other studies have focused on data-driven techniques for process decomposition. For example, Ge et al. \cite{Ge2013} proposed a method to create monitoring blocks based on process variables' contributions to principal components obtained from PCA. This approach groups variables according to their influence on the principal components derived from the PCA decomposition of the entire process. Following a different approach, Deng et al. \cite{DengZiqing2021} proposed an algorithm based on joint mutual information to generate monitoring blocks, capturing the dependency among multiple variables. Using projective dictionary pair learning, they then created local monitoring models for each block. Yin et al. \cite{Yin2022} proposed constructing an undirected graph to represent relationships between process variables using mutual information. Then, a modified fast-unfolding algorithm was applied to detect communities within this structure and construct monitoring blocks. Similarly, in their recent work, Yin et al. \cite{Yin2024} proposed using the Girvan-Newman algorithm \cite{Girvan2002} for process decomposition and incorporating probabilistic deep learning models in the monitoring blocks. While the monitoring approach by Yin et al. \cite{Yin2024} has demonstrated improved performance compared to linear methods, it requires careful hyperparameter tuning to ensure robustness, which poses scalability challenges for more extensive processes.

These studies suggest that inter- and intra-subsystem interactions are essential in improving fault detection performance in decentralized monitoring systems. However, challenges still need to be addressed, primarily when the sensors are unevenly distributed across the plant. For instance, in knowledge-based process decomposition, there is no consensus on optimally structuring monitoring blocks in such cases: some methods retain blocks with few measurements, while others merge them, even if that implies grouping non-contiguous process units. This lack of standardization makes it difficult to determine if variations in FDI performance are due to process decomposition or the monitoring model itself (e.g., PCA, PLS, CCA), as different experts may create different monitoring blocks for the same plant. Moreover, existing knowledge-based process decomposition methods often overlook valuable information from the plant's piping and instrumentation diagram (P\&IDs). Control pairs, which are variables with solid correlations and established causal relationships, provide insights that could enhance the FDI performance of monitoring blocks.

While data-driven process decomposition methods leverage the abundance of available plant data and can achieve automatic decomposition without requiring extensive expert knowledge, the complexity of these techniques is creating a gap between the training of the average control engineer and the skills needed to implement them efficiently. For example, generating graph representations based on statistical properties is less intuitive than analyzing a PFD and may not be readily accessible to practitioners without specialized expertise. This complexity can hinder the adoption of advanced data-driven monitoring systems in industry, where control engineers may need more specialized training in advanced statistical methods or machine learning algorithms. Moreover, existing knowledge-based decomposition methods have limitations, such as the need for more standardization and the potential to overlook valuable information from plant control schemes. Therefore, there is a pressing need for process decomposition methods that are both effective and accessible to the average control engineer.

In order to bridge the gap between complex data-driven methods and accessible knowledge-based approaches, this study proposes a structured knowledge-based process decomposition algorithm for decentralized process monitoring combined with full Principal Component Analysis (f-PCA). Our method integrates control loop structure information with topological data from the PFD, all within the skill set of an average control engineer. Our primary objective was to develop a process decomposition approach that automatically generates monitoring blocks without requiring additional correlation or mutual information data among process variables, making it suitable for plants with uneven sensor distribution. This allows control engineers to implement robust monitoring systems without additional specialized training or reliance on external experts.

The results of testing on the Tennessee Eastman Process (TEP) benchmark show that the proposed methodology enables straightforward identification of fault locations within the plant and analysis of fault propagation through the plant. Furthermore, the monitoring results demonstrate that the fault detection and false alarm rates are comparable or better than the results achieved with more involved, complex centralized or distributed methods for fault detection.

The main contributions of this work are summarized as follows:
\begin{itemize}
    \item A systematic process decomposition framework that integrates PFD and control scheme data to form well-structured monitoring blocks for decentralized process monitoring.
    \item A set of heuristic rules for merging monitoring blocks with few variables that preserve contiguity and prevent module overload.
    \item A reformulation of the contributions map tailored for distributed monitoring systems to ensure consistent block comparison on a unified scale. These reformulated contributions streamline the analysis of fault propagation through the plant.
\end{itemize}

The remainder of this paper is organized as follows: Section \ref{sec:met} details the workflow and structure of the FDI framework, including the proposed process decomposition technique and the reformulation of contribution plots. Section \ref{sec:case} presents a case study evaluating the performance of the proposed FDI framework on the Tennessee Eastman Process benchmark. Finally, Section \ref{sec:conc} provides conclusions and insights into future work.

\section{Methodology} \label{sec:met}
This section describes the methodology for the proposed decentralized monitoring framework. First, we introduce the novel process decomposition technique. Next, we describe how fault detection is performed within each monitoring block using a full PCA model. Then, we outline how information from all blocks is combined with a Bayesian fusion rule to produce a global anomaly score. Finally, we detail the fault identification process, which uses an enhanced version of the contributions map designed to improve performance in decentralized process monitoring.

\subsection{Process Decomposition}
Process decomposition based on topological attributes has often utilized community detection algorithms that maximize modularity. While this approach effectively identifies non-overlapping communities in undirected graphs, the problem becomes more complex when considering information flow, overlapping communities, and information distribution. In industrial complexes, unit operations interact through material and energy flows or control signals. The proposed knowledge-based algorithm considers these interactions along with sensor distribution within the plant.

Before detailing the proposed algorithm, we first introduce the Measurement Allocation Ratio (MAR) of a subgraph \( S_i \), defined as:
\begin{equation}
\label{eq:MAR}
\begin{aligned}
\text{MAR}(S_i) = \frac{M(S_i)}{\sum_{j} M(S_j)},
\end{aligned}
\end{equation}
where \( M(S_i) \) is the number of measurements in \( S_i \), and the denominator represents the total measurements across all subgraphs.

The proposed algorithm comprises two phases. In the first phase, an initial set of monitoring blocks \( \mathcal{B} = \{ B_i \mid i = 1, 2, \dots, n \} \) is generated based on the measurement allocation in the plant. In the second phase, this initial set of monitoring blocks is refined using information from the process control loop pairs \( \mathcal{Q} = \{(CV_1, MV_1), \dots, (CV_n, MV_n)\} \). A satisfactory process decomposition can be achieved by executing only the first phase of the proposed algorithm, referred to as MAR-based process decomposition. If both phases are executed, the algorithm is called control-aware MAR-based process decomposition.

\subsubsection{Phase One: MAR-based decomposition algorithm}
This algorithm iteratively merges communities (subgraphs) within a directed graph based on a measurement allocation threshold. The steps are detailed below:

\begin{enumerate}
    \item \textbf{Graph Initialization:} The process flow diagram is transformed into a directed graph by adding extra blocks where necessary (e.g., mixers and splitters), similar to the generation of state task networks for batch processes \cite{Maravelias2003}. This results in a directed graph \( G(V,E) \), where \( V = V_S \cup V_U \), with \( V_S \) representing stream nodes and \( V_U \) representing unit operation nodes. The edges \( E \) model the directional dependencies between nodes.

    \item \textbf{Identifying Initial Subgraphs:} Each initial subgraph \( S_i \subset G \) is centered around a unit operation node \( u \in V_U \) and includes all directly connected stream nodes \( s \in V_S \), such that the connected subgraph \( S_i \) is given by: 

    \begin{equation}
    \label{eq:subS} 
    S_i = \{ u \} \cup \{ s \in V_S \mid (s, u) \in E \lor (u, s) \in E \}
    \end{equation}

    Then all the initial subgraphs define the set \( \mathcal{S}_k = \{S_i \mid i = 1, 2, \dots, n \} \).

    \item \textbf{Calculation of Measurement Allocation Ratio (MAR):}    
    Calculate the MAR for each subgraph \( S_i \in \mathcal{S}_k \), using Eq. \ref{eq:MAR}. Subgraphs with a MAR value below a specified threshold \( \delta \) are selected for merging.

    \item \textbf{Subgraph Merging:} 
    \begin{enumerate}
        \item Sort subgraphs in ascending order of MAR values, so subgraphs with fewer measurements are processed first. Let \( \mathcal{P} = \{S_1, S_2, \dots, S_m\} \) represent the merging pool after sorting, where \( \text{MAR}(S_i) \leq \text{MAR}(S_{i+1}) \) for \( i = 1, \dots, m-1 \).
         
        \item For each subgraph \( S_i \in \mathcal{P} \), we find the set of neighboring subgraphs \( \mathcal{N} \), where \( S_j \in \mathcal{N} \) if there exists a directed edge \( (s_i, u_j) \in E \) with \( s_i \in S_i \) and \( u_j \in S_j \). Note that \( S_j \in \mathcal{N} \) can be within or outside \( \mathcal{P} \).
        
        \item Then we find the best merge \( S_{j}^{\text{best}} \) for each \( S_i \) as:

           \begin{equation}
           \label{eq:BestS} 
           S_{j}^{\text{best}} = \underset{S_j \in \mathcal{N}}{\operatorname{argmin}} \ \text{MAR}(S_j).
           \end{equation}
           
         \item Then we merge \( S_i \) with its respective \( S_{j}^{\text{best}} \) following the sequence in \( \mathcal{P} \). If a \( S_i \) or \( S_{j}^{\text{best}} \) has already merged in the current iteration, it is not considered for further merging to ensure each subgraph merges only once per iteration. This restriction maintains an orderly merging process across iterations.
         
         \item After the merging pool is exhausted, \( \mathcal{S}_k \) is updated to include the new subgraphs.
         
    \end{enumerate}

    \item \textbf{Iteration Until Convergence:} Repeat Steps 3 and 4. In each iteration, recalculate MAR values for the new subgraphs in \( \mathcal{S}_{k+1} \) and continue merging until the threshold criteria are met, with no further merges possible. 

    \item \textbf{Output of Final Subgraphs:} After the final iteration, the algorithm concludes by constructing the process monitoring blocks \( B_i \), which encompass all variables within their respective subgraphs \( S_i \in \mathcal{S}_{\text{final}} \).
\end{enumerate}

\subsubsection{Phase Two: Monitoring Blocks Refinement}
In the second phase, the monitoring blocks obtained in the first phase are refined by incorporating control loop information. The steps are detailed below:

\begin{enumerate}
    
    \item \textbf{Retrieving Control Loop Information:} From the control scheme, we define a set \( \mathcal{Q} = \{(CV_1, MV_1), \dots, (CV_n, MV_n)\} \), where each pair \( (CV_i, MV_i) \) consists of a controlled variable \( CV_i \) and its corresponding manipulated variable \( MV_i \). For each control loop pair \( (CV_i, MV_i) \in \mathcal{Q} \), let:
    
    \begin{equation}
    \label{eq:BeCV} 
    B_{CV} = \{B_j \mid CV_i \in B_j\}
    \end{equation}
    
    \begin{equation}
    \label{eq:BeMV} 
    B_{MV} = \{B_k \mid MV_i \in B_k\}.
    \end{equation}
    
    Here, \( B_{CV} \) represents the block containing \( CV_i \), and \( B_{MV} \) represents the block containing \( MV_i \).
        
    \item \textbf{Ensuring Control Loop Cohesion:} For each pair \( (CV_i, MV_i) \in \mathcal{Q} \), if \( B_{CV} \neq B_{MV} \), then reassign \( MV_i \) to the block \( B_{CV} \) to ensure both control loop variables belong to the same monitoring block. This reassignment is performed as:
    \[
    B_{CV} := B_{CV} \cup \{MV_i\}.
    \]
    This step ensures that \( MV_i \) is added to the block containing \( CV_i \), preserving the control relationship within a single monitoring block.

    \item \textbf{Completion of the Phase Two:} Repeat this refinement process for all control loop pairs in \( \mathcal{Q} \) until each pair has been evaluated and adjusted as necessary. The second phase concludes once all pairs have been reviewed, yielding a set of refined monitoring blocks where each control loop pair \( (CV_i, MV_i) \in \mathcal{Q} \) is contained within a single block.
    
\end{enumerate}

\subsection{Block Level Fault Detection}
As a result of the process decomposition, we obtain a set \( \mathcal{B} = \{ B_i \mid i = 1, 2, \dots, n \} \) of monitoring blocks. For each monitoring block $B_i \in \mathcal{B}$, a PCA model is trained using the normal operation data. The number of principal components to retain in a PCA model is typically determined by parallel analysis \cite{Horn1965} or cross-validation \cite{Wold1978}. However, Tamura et al. \cite{Tamura2007} showed that the sensitivity of Hotelling's $T^2$ increases as more principal components are included in the PCA model. Additionally, they found that some faults can only be detected through components that account for relatively little variance. In this work, full PCA is employed in each monitoring block to improve fault detection robustness.

Let us consider $\boldsymbol{X}_b$ as the data matrix corresponding to the block $B_i \in \mathcal{B}$, then through PCA $\boldsymbol{X}_{b}$ is decomposed into:

\begin{equation}
\label{eq:PCA}
\begin{aligned}
\boldsymbol{X}_b = \boldsymbol{T}_b \boldsymbol{P}_b^T + \boldsymbol{E}_b ,
\end{aligned}
\end{equation}

where $ \boldsymbol{T}_b \in \mathbb{R}^{m \times k} $ is the scores matrix, $ \boldsymbol{P}_b \in \mathbb{R}^{n \times k} $ is the loadings matrix, and  $ \boldsymbol{E}_b \in \mathbb{R}^{m \times n}$ is the residual matrix. When full PCA is used , the process is monitored using Hotelling’s $T^2$. From each row in $ \boldsymbol{T}_b$, Hotelling’s $T^2$ statistic is calculated as:

\begin{equation}
\label{eq:T2}
\begin{aligned}
T_i^2 = \boldsymbol{t}_i^T \boldsymbol{\Lambda} \boldsymbol{t}_i, 
\end{aligned}
\end{equation}

where $\boldsymbol{\Lambda} \in \mathbb{R}^{k \times k}$ is the diagonal matrix containing the variance of each column of $\boldsymbol{T}_b$. Once an observation’s \( T_i^2 \) exceeds the local control limit, the observation is labeled as abnormal. The control limit for Hotelling’s \( T^2 \) in the PCA model for $B_i \in \mathcal{B}$ is calculated as:

\begin{equation}
\label{eq:T2lim}
\begin{aligned}
T_{b,\text{lim}}^2 = \frac{(N^2 - 1)p}{N(N - p)} F_{\alpha, p, N-p} ,
\end{aligned}
\end{equation}

where $p$ is the number of variables, $N$  is the number of observations, and  $F_{\alpha, p, N-p}$ is the critical value from the F-distribution with $p$ and $N-p$ degrees of freedom at a significance level $\alpha$.

\subsection{Bayesian Fusion}
In distributed process monitoring, it is essential to aggregate anomaly scores from multiple monitoring blocks into a unified global anomaly score. Bayesian inference facilitates this by converting the monitoring results into fault probabilities, ensuring comparability across all monitoring blocks on a consistent scale. This process is given by:

\begin{equation}
\label{eq:PFXnew}
\begin{aligned}
P_{T^2}^{b}(F \mid \boldsymbol{x}_{\text{i}}) = \frac{P_{T^2}^{b}(\boldsymbol{x}_{\text{i}} \mid F) P_{T^2}^{b}(F)}{P_{T^2}^{b}(\boldsymbol{x}_{\text{i}})}
\end{aligned}
\end{equation}

\begin{equation}
\label{eq:PXnew}
\begin{aligned}
P_{T^2}^{b}(\boldsymbol{x}_{\text{i}}) = P_{T^2}^{b}(\boldsymbol{x}_{\text{i}} \mid F) \cdot P_{T^2}^{b}(F) + P_{T^2}^{b}(\boldsymbol{x}_{\text{i}} \mid N) \cdot P_{T^2}^{b}(N),
\end{aligned}
\end{equation}

where $P_{T^2}^{b}(N)$ and $P_{T^2}^{b}(F)$ are the prior probabilities of the faulty and normal conditions, respectively. These are defined in terms of the significance level $\alpha$ as:

\begin{equation}
\label{eq:PN}
\begin{aligned}
P_{T^2}^{b}(N) = 1 - \alpha
\end{aligned}
\end{equation}

\begin{equation}
\label{eq:PF}
\begin{aligned}
P_{T^2}^{b}(F) = \alpha.
\end{aligned}
\end{equation}

In equations \ref{eq:PN} and \ref{eq:PF}, $\alpha$ is the significance level at which $T_{b,\text{lim}}^2$ is calculated. For a new data sample, the likelihoods \( P_{T^2}^{b}(\boldsymbol{x}_{\text{i}} \mid N) \) and \( P_{T^2}^{b}(\boldsymbol{x}_{\text{i}} \mid F) \), representing normal and faulty conditions respectively, are computed as follows \cite{Ge2010}:

\begin{equation}
\label{eq:PXnewN}
\begin{aligned}
P_{T^2}^{b}(\boldsymbol{x}_{\text{i}} \mid N) = \exp \left( -\frac{T_{b, \text{i}}^2}{T_{b, \text{lim}}^2} \right)
\end{aligned}
\end{equation}

\begin{equation}
\label{eq:PXnewF}
\begin{aligned}
P_{T^2}^{b}(\boldsymbol{x}_{\text{i}} \mid F) = \exp \left( -\frac{T_{b, \text{lim}}^2}{T_{b, \text{i}}^2} \right)
\end{aligned}
\end{equation}

After calculating fault probabilities for each monitoring block, the global anomaly score is obtained by aggregating the posterior probabilities using the Bayesian Inference Criterion (BIC) \cite{Ge2010}. The BIC weighs the relative contributions of each monitoring block and is defined as:

\begin{equation}
\label{eq:BIC}
\begin{aligned}
\text{BIC}_{T^2}(\boldsymbol{x}_{\text{i}}) = \sum\limits_{b \in \mathcal{B}} \frac{P_{T^2}^{b}(\boldsymbol{x}_{\text{i}} \mid F) \cdot P_{T^2}^{b}(F \mid \boldsymbol{x}_{\text{i}})}{\sum\limits_{b \in \mathcal{B}} P_{T^2}^{b}(\boldsymbol{x}_{\text{i}} \mid F)}.
\end{aligned}
\end{equation}

This approach ensures that the fault probabilities are effectively combined across monitoring blocks, enabling a reliable global anomaly detection framework.

\begin{figure}[t] 
    \centering
    \includegraphics[width=0.45\textwidth]{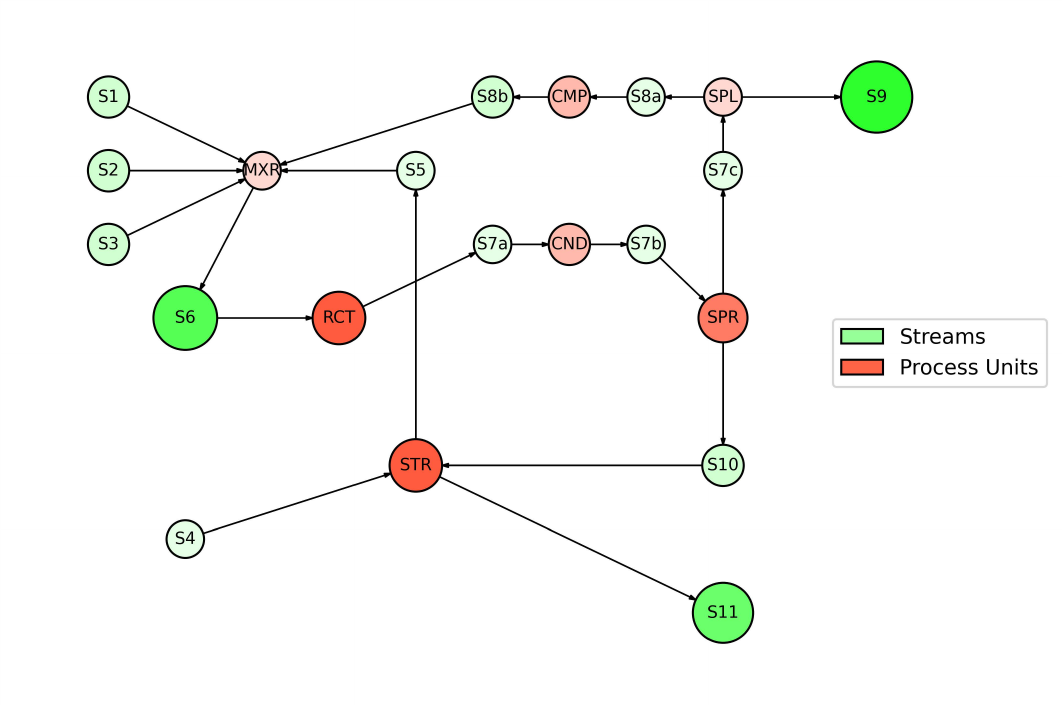} 
    \caption{TEP digraph representation.}
    \label{fig:DiGraph}
\end{figure}

\subsection{Fault Identification}
In an FDI system, once a fault is detected, an analysis of each original variable's contribution to the anomaly score is conducted to determine the variables most strongly linked to the fault. A contributions map is commonly used for this purpose, as it provides a clear graphical representation of each variable's contribution to the anomaly score. In PCA, different methods exist to calculate the contributions when the process is monitored using Hotelling's $T^2$  \cite{Miller1998, Westerhuis2000, Kourti1996}. The main difference among these techniques lies in which terms of the linear combination defining a score $t_i$ are considered as contributions. In order to help control engineers better understand how faults propagate across space and time, contributions are visualized in 2D maps, where observations are stacked into a single color representation \cite{Zhu2014}. While effective in centralized monitoring systems, this approach presents some challenges in distributed systems:

\begin{itemize}
    \item Contributions often vary in scale across different monitoring blocks, making comparison harder. 
    \item Dominant variables with rapidly growing contributions can obscure those whose effects take longer to emerge.
\end{itemize}

To address these limitations, we propose a clipped contribution ratio map. Instead of plotting raw contributions, we scale each contribution by its respective block $T^2$ control limit and clip the ratio at one. This ensures that once a variable's contribution exceeds the control limit, further escalation does not obscure the contributions of other variables. By preventing dominant variables from overwhelming the visualization, this method maintains a balanced view of all relevant variables, improving interpretability as shown in Section \ref{sec:case}.

The contributions of each process variable $x^b_j$ in $b \in \mathcal{B}$ to each $ {t^b}_i$ in the score vector ${\boldsymbol{t}^b}$ is calculated as:

\begin{equation}
\label{eq:cont}
\text{cont}(t^b_i, x^b_j) = \max\left(0, \frac{t^b_i P^b_{ij} x^b_j}{T_{b, \text{lim}}^2}\right)
\end{equation}

where $P^b_{ij}$ is the $ij$-th element of the loading matrix $\boldsymbol{P}_b$. Thus, the total contribution of the process variable $x^b_j$ to the $ T^2_b$ is calculated as:

\begin{equation}
\label{eq:totcont}
\text{CONT}(x^b_j) = \min\left(1.0, \sum_i \text{cont}(t^b_i, x^b_j)\right).
\end{equation}

\begin{figure}[t] 
    \centering
    \includegraphics[width=0.48\textwidth]{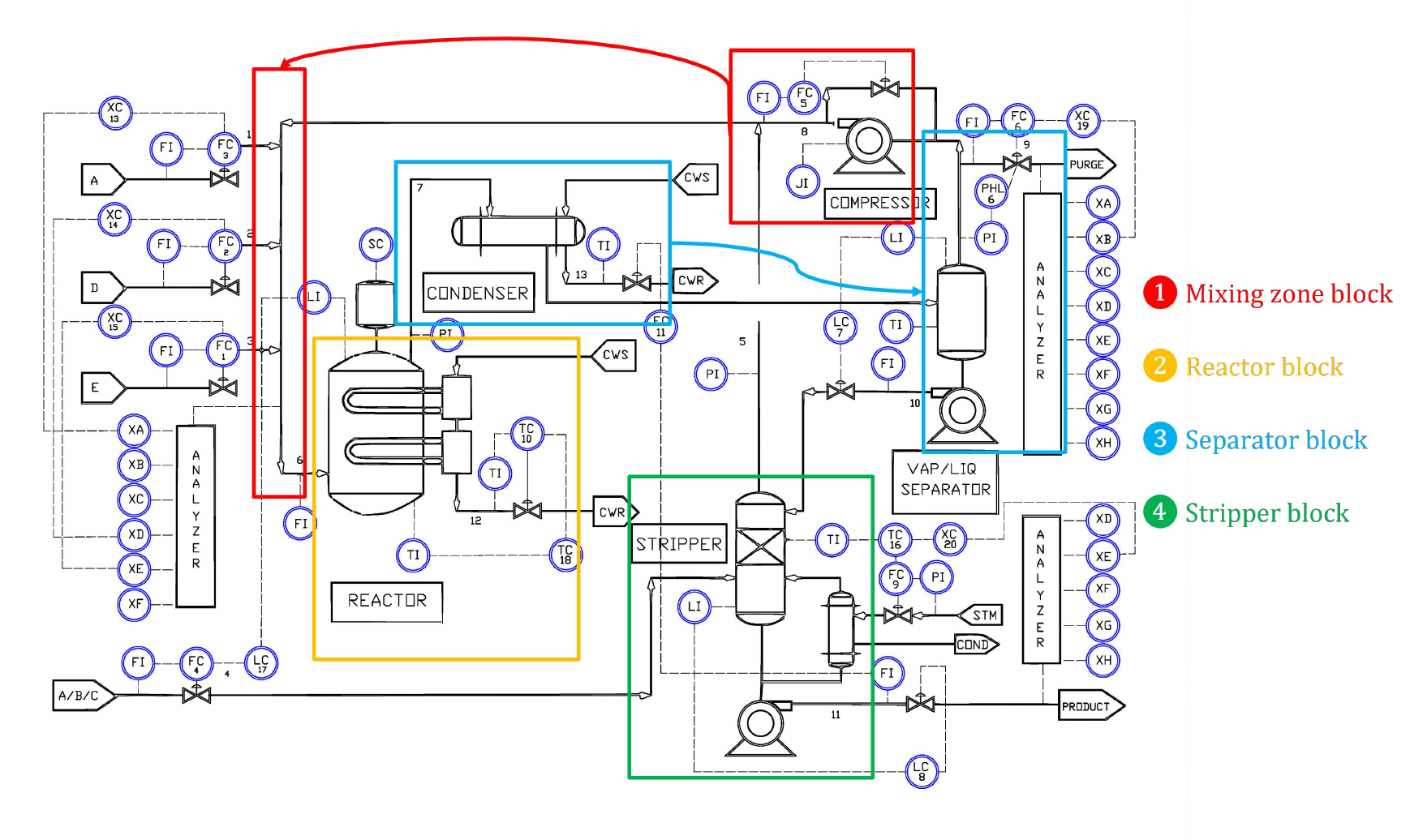} 
    \caption{TEP monitoring blocks derived from the MAR-based decomposition algorithm.}
    \label{fig:Modules}
\end{figure}

\section{Case Study: The Tennessee Eastman Process} \label{sec:case}

\subsection{Problem Description}

Originally proposed by Downs \& Vogel \cite{Downs1993}, the Tennessee Eastman Process (TEP) involves eight chemical components and five main process units in a reactor/separator/recycle configuration. The system features two primary exothermic reactions (A + C + D → G, A + C + E → H) and two secondary byproduct reactions (A + E → F, 3D → 2F), all occurring in the presence of an inert gas (B). There are various TEP datasets available, mainly differing in the control scheme applied to the process \cite{McAvoy1994, Lyman1995,Ricker1996}. In this work, we utilize the dataset presented in \cite{Russell2000}, which can be downloaded from a Git repository \cite{Russell2000b}.

\begin{table}[t]
\centering
\caption{Comparison of the Proposed Method and Existing Decentralized Methods Based on FAR (\%) for Fault 0 and FDR (\%) for Faults 1–21}

\label{tab:table1}
\resizebox{\textwidth}{!}{%
\begin{tabular}{ccccccc}
\hline
\textbf{Fault} & \textbf{Control-Aware} & \textbf{MAR-based} & \textbf{Louvain-based} & \textbf{Fully-connected} & \textbf{Partly-connected} & \textbf{Distributed} \\
&\textbf{MAR-based PCA}&\textbf{PCA} & \textbf{PCA}& \textbf{DCCA} \cite{Chen2019} & \textbf{DCCA} \cite{Peng2020} &  \textbf{PCA} \cite{Ge2013} \\
 & \( \text{BIC}_{T^2} \) & \( \text{BIC}_{T^2} \) & \( \text{BIC}_{T^2} \) & \( T^2_{local} \) & \( T^2_{local} \) & \( T^2_{local} \) \\
\hline
0 & 3.23 & \textbf{3.13} & 3.65 & 4.90 & 5.10 & - \\
\hline
3 & 3.88 & \textbf{3.75} & 5.63 & - & - & 7.40 \\
9 & 3.75 & \textbf{3.50} & 4.75 & - & - & 8.70 \\
15 & 13.50 & 12.25 & 14.13 & - & - & \textbf{7.70} \\
\hline
4 & \textbf{100.00} & \textbf{100.00} & \textbf{100.00} & \textbf{100.00} & 99.75 & \textbf{100.00} \\
5 & \textbf{100.00} & 28.13 & 29.50 & 43.36 & 42.38 & \textbf{100.00} \\
7 & \textbf{100.00} & \textbf{100.00} & \textbf{100.00} & 54.64 & 53.63 & 88.20 \\
\hline
1 & 99.63 & 99.63 & 99.75 & 99.62 & 99.38 & \textbf{99.90} \\
2 & 98.63 & 98.50 & 98.75 & \textbf{99.49} & 99.13 & 97.70 \\
6 & \textbf{100.00} & \textbf{100.00} & \textbf{100.00} & 99.75 & 99.75 & \textbf{100.00} \\
8 & \textbf{97.75} & 97.63 & \textbf{97.75} & 92.48 & 91.75 & 95.90 \\
10 & 86.38 & 83.00 & 84.50 & \textbf{90.85} & 90.50 & 53.40 \\
11 & 78.75 & 78.75 & 79.88 & 73.43 & 73.38 & \textbf{83.20} \\
12 & \textbf{99.75} & 99.50 & 99.50 & 94.61 & 94.63 & 98.20 \\
13 & 95.25 & 95.25 & 95.38 & 86.87 & 86.13 & \textbf{95.70} \\
14 & \textbf{100.00} & \textbf{100.00} & \textbf{100.00} & 93.36 & 90.25 & 99.90 \\
16 & \textbf{92.00} & 91.50 & 52.75 & 26.24 & 72.38 & 48.10 \\
17 & 95.50 & 95.13 & 95.25 & 88.85 & 90.88 & \textbf{97.40} \\
18 & 91.00 & 90.88 & 91.50 & \textbf{93.23} & 92.88 & 91.00 \\
19 & 58.00 & 57.25 & 37.88 & 63.86 & \textbf{89.25} & 45.20 \\
20 & \textbf{79.75} & 60.63 & 63.63 & 72.81 & 72.88 & 65.20 \\
21 & \textbf{55.38} & 48.75 & 51.88 & 49.51 & 48.50 & 50.00 \\
\hline
\end{tabular}%
}
\end{table}

The repository provides training and testing datasets for 21 faults, as well as data representing normal process operation. Each testing dataset simulates 48 hours of plant operation, containing 41 process measurements (XMEAS(1)-XMEAS(41)) and 11 manipulated variables (XMV(1)-XMV(11)) sampled every 3 minutes, totaling 960 samples. The first 8 hours (160 samples) represent normal operation, while the remaining 40 hours (800 samples) capture the control system's response to disturbances (IDV (1)-IDV (21)). Sun et al. \cite{Sun2020} proposed classifying TEP faults into three categories: controllable faults, back-to-control faults, and uncontrollable faults. Controllable faults can be mitigated by the control system without pushing any measured or manipulated variables out of their normal operating ranges. Back-to-control faults occur when the control system restores the measured variables to the normal state, but one or more manipulated variables move outside their normal range. Uncontrollable faults are those in which neither the measured nor the manipulated variables can be returned to their normal operating regions.

The three process decomposition algorithms outlined above were applied to the TEP digraph, resulting in five monitoring blocks with the Louvain algorithm, and four blocks with both the partial and full MAR-based algorithms. Fig. \ref{fig:Modules} illustrates the monitoring blocks derived from the MAR-based algorithm.

\subsection{Monitoring Blocks}
Initially, a graph representation of the TEP was generated from its process flow diagram. This involved adding a mixer to join streams S1, S2, S3, S5 and S8, and a splitter to divide the separator's effluent into purge and recycle. The resulting directed graph representation of the TEP is shown in Fig. \ref{fig:DiGraph}, where the size of the circles is proportional to the number of measurements available in each node. The process decomposition necessary for implementing the distributed process monitoring system was conducted using three approaches: the Louvain algorithm \cite{Fu2008}, the MAR-based algorithm, and the control-aware MAR-based algorithm (see Section \ref{sec:met}) with $\delta = 0.15 $. The MAR-based techniques resulted in the following unit operation-based decomposition: 1) Mixing Zone/Compressor, 2) Reactor, 3) Separator/Condenser, and 4) Stripper.

The decomposition presented in Chen et al. \cite{Chen2019}, later reused in Peng et al. \cite{Peng2020}, consists of the following blocks: 1) Reactor/Mixing Zone, 2) Separator, 3) Condenser/Compressor, and 4) Stripper. In their method, each process variable is assigned to a unique monitoring block, and blocks can include non-contiguous process units.

Ensuring that blocks represent contiguous areas of the process enhances fault isolation during the detection step and simplifies the visualization of fault propagation paths on contribution maps. For example, in the TEP, an alarm in a Condenser/Compressor monitoring block does not immediately indicate whether the fault originated in the condenser or the compressor without further analysis of the separator's alarm status or the block's contribution map. In contrast, with contiguous process units grouped in a monitoring block, such as Separator/Condenser, narrowing down the fault's location requires no cross-block analysis, facilitating isolation at the sector level. Additionally, changes in the condenser variables have a more direct and immediate effect on the separator variables than on the compressor variables.

In the decomposition presented in \cite{Ge2013}, the TEP is monitored using 15 blocks. In this approach, process variables can belong to multiple blocks, increasing the potential to capture a wide range of local patterns. Since there are no topological restrictions on structuring these blocks, faults cannot be isolated at the plant sector level during the detection step, and the alarm sequence analysis may not clearly reveal a fault propagation path. Additionally, tracking the behavior of five process units across 15 blocks is more challenging for the control room operator.

\subsection{Fault Detection}
A full PCA model was constructed for each $B_i \in \mathcal{B}$, where \( \mathcal{B} \) represents the set of monitoring blocks derived from a specific decomposition technique. In order to facilitate the FDI analysis, Hotelling's \( T^2_b \) profiles were expressed in terms of probabilities \( P_{T^2}^{b} (F | x_{\text{i}})\), with the threshold \(T_{b, \text{lim}}^2\) determined at a significance level \( \alpha = 0.01 \). The local posterior probabilities were then combined using Eq. \ref{eq:BIC} to monitor the entire system. Due to noise in process measurements, normal observations may sometimes exceed the detection threshold. To accurately detect faulty behavior, it is standard practice to declare a fault only after observing a sequence of \( N \) consecutive readings above the threshold \cite{Russell2000}. In this study, faulty operation is confirmed once seven consecutive observations exceed the detection threshold.

\begin{table}[h]
\centering
\caption{Comparison of the Proposed Method and Existing Centralized Methods Based on FAR (\%) for Fault 0 and FDR (\%) for Faults 1–21}
\label{tab:table2}
\resizebox{\textwidth}{!}{%
\begin{tabular}{c c c cc cc cc c}
\hline
\textbf{Fault} & \textbf{CMAR-PCA} & \textbf{f-PCA} \cite{Sun2020} & \multicolumn{2}{c}{\textbf{KPCA} \cite{Cacciarelli2022}} & \multicolumn{2}{c}{\textbf{AE} \cite{Cacciarelli2022}} & \multicolumn{2}{c}{\textbf{OAE} \cite{Cacciarelli2022}} &\textbf{BRNN} \cite{Sun2020} \\ 
               & \( \text{BIC}_{T^2} \) & \( T^2 \)&\( T^2 \) & \( SPE \)&\( T^2 \) &\( SPE \)&\( T^2 \) & \( SPE \)& \( D^2 \)     \\ 

\hline
0 & 5.00 & 5.00 & 5.10 & 5.00 & 4.90 & 5.10 & 5.90 & 5.20 & \textbf{4.75} \\
\hline
3 & 6.38 & 19.75 & 5.50 & 5.20 & 5.30 & 5.30 & 6.60 & 5.70 & \textbf{5.00} \\ 
9 & \textbf{4.88} & 15.25 & 5.70 & 5.30 & 5.50 & 5.40 & 6.90 & 6.00 & 5.00 \\ 
15 & 16.25 & 26.87 & 6.40 & \textbf{5.40} & 6.20 & 5.60 & 7.50 & 7.00 & 7.12 \\ 
\hline
4 & \textbf{100.00} & \textbf{100.00} & 67.80 & \textbf{100.00} & 50.80 & \textbf{100.00} &\textbf{100.00}& 99.60 & \textbf{100.00} \\ 
5 & \textbf{100.00} &\textbf{100.00} & 30.80 & 20.50 & 32.20 & 27.20 & \textbf{100.00} & \textbf{100.00} & \textbf{100.00}\\ 
7 & \textbf{100.00} & \textbf{100.00} & \textbf{100.00} & \textbf{100.00} & \textbf{100.00} & \textbf{100.00} & \textbf{100.00} & \textbf{100.00} & \textbf{100.00} \\ 
\hline
1 & 99.63 & \textbf{100.00} & 99.80 & 99.40 & 99.30 & 99.80 & 99.80 & 99.70 & 99.75 \\ 
2 & 98.63 & \textbf{99.12} & 98.70 & 96.20 & 98.60 & 97.40 & 98.80 & 98.30 & 99.00 \\ 
6 & \textbf{100.00} & \textbf{100.00} & 99.40 & \textbf{100.00} & 99.90 & \textbf{100.00} & \textbf{100.00} & \textbf{100.00} & \textbf{100.00} \\ 
8 & 97.75 & \textbf{98.25} & 97.80 & 85.40 & 96.70 & 92.20 & 97.90 & 97.60 & 98.12 \\ 
10 & 87.13 & \textbf{93.50} & 41.00 & 33.70 & 35.10 & 43.00 & 80.70 & 90.70 & 87.38 \\ 
11 & 79.25 & \textbf{87.25} & 59.40 & 78.80 & 54.20 & 84.70 & 86.00 & 74.00 & 74.75 \\ 
12 & 99.75 & \textbf{100.00} & 98.60 & 93.80 & 98.50 & 95.10 & 99.10 & 99.10 & 99.75 \\ 
13 & 95.25 & \textbf{95.75} & 95.00 & 90.00 & 94.30 & 93.10 & 95.40 & 94.90 & \textbf{95.75}\\ 
14 & \textbf{100.00} & \textbf{100.00} & 99.90 & 99.90 & 99.80 & \textbf{100.00} & \textbf{100.00} & \textbf{100.00} & \textbf{100.00} \\ 
16 & 93.00 & \textbf{95.50} & 21.90 & 29.80 & 18.60 & 38.00 & 82.00 & 94.00 & 90.38 \\ 
17 & 95.50 & \textbf{97.75} & 84.60 & 92.10 & 79.10 & 94.10 & 94.90 & 92.00 & 96.13 \\ 
18 & 91.13 & 91.50 & 93.30 & 94.20 & 93.50 & 94.20 & \textbf{94.50} & 94.00 & 90.63 \\ 
19 & 62.63 & \textbf{96.00} & 27.10 & 32.80 & 34.50 & 27.40 & 84.90 & 91.60 & 88.25 \\ 
20 & 81.00 &\textbf{92.13} & 50.10 & 52.40 & 38.30 & 59.20 & 75.10 & 80.00 & 78.63 \\ 
21 & 56.25 &\textbf{61.62}& - & - & - & - & - & - & 48.00 \\ 
\hline

\end{tabular}
}
\end{table}

The performance of the monitoring systems is evaluated using the Fault Detection Rate (FDR), and the False Alarm Rate (FAR). The FDR is defined as follows:

\begin{equation}
\text{FDR} = \frac{\text{TP}}{\text{TP} + \text{FN}}
\end{equation}

where $\text{TP}$ represents the number of samples flagged with an alarm after the fault is introduced, and $\text{FN}$ is the number of faulty samples labeled as normal. The FAR is defined as:

\begin{equation} \text{FAR} = \frac{\text{FP}}{\text{FP} + \text{TN}} \end{equation}

where $\text{FP}$ is the number of normal samples incorrectly flagged as faults, and $\text{TN}$ is the number of normal samples correctly identified. Robust fault detection requires high FDR and low FAR to minimize missed detections and false alarms.

The fault detection results are summarized in Tables \ref{tab:table1} and \ref{tab:table2}. Table \ref{tab:table1} compares various distributed monitoring processes, including those proposed in this study, while Table \ref{tab:table2} contrasts the Control-Aware MAR-based PCA with centralized approaches, including nonlinear techniques.

Fault detection is typically performed using a significance level of 0.01 or 0.05, which poses challenges for comparison since researchers often report results at a single level. In alignment with Russell et al. \cite{Russell2000}, we recalculated the detection threshold for $\text{BIC}_{T^2}$ to achieve a 5\% false alarm rate (FAR) on the validation dataset. This adjustment facilitates a fair comparison of our approach with the methods depicted in Sun et. al \cite{Sun2020} and Cacciarelli et. al \cite{Cacciarelli2022} (see Table \ref{tab:table2}).

The distributed methods introduced in this work, including Control-Aware MAR-based PCA (CMAR-PCA) and MAR-based PCA (MAR-PCA), demonstrate excellent fault detection performance. They achieve 100\% detection rates for faults 4, 5, 6, 7, and 14, and over 97\% detection rates for faults 1, 2, 8, and 12. Notably, the FDR for fault 7 significantly outperforms DCCA variants. For faults where the FDR is slightly lower (faults 1, 2, 13, 17, and 18), the difference is minor, averaging about 2\%. Faults 10 and 11 show a slightly larger difference, averaging 5\%. However, the partly connected DCCA method performs better for fault 19. The CMAR-PCA approach uniquely integrates process knowledge by structuring the merging of blocks with low information based on the plant's measurement allocation. This ensures effective integration of control loops and monitoring blocks with limited sensors, addressing the challenges of large systems where extensive correlation analysis is not feasible. The small performance gap compared to techniques that ignore control loops or structured merging is therefore acceptable.

Considering control loop information in the process decomposition significantly enhance fault detection rates for specific faults, as demonstrated by our results. For example, the detection rate of Fault 5 improves because the control aware MAR-based algorithm assigns the condenser steam flow to the stripper monitoring block. Similarly, Fault 7 benefits from the inclusion of stream 4’s flow in the reactor monitoring block. The approach proposed by Ge et al. \cite{Ge2013} may also capture control relationships indirectly through principal component contribution-based clustering, but it is more complex than our method.

In the case of Fault 16, the monitoring block for the stripper defined in this work is identical to that proposed by Chen et. al \cite{Chen2019}. However, we believe the increase in the fault detection rate is due to the number of principal components retained in our study. Since the components that reveal this fault likely account for a small portion of the total variance, using full PCA allows us to better represent the stripper column operation \cite{Tamura2007}. On the other hand, Ge et al. \cite{Ge2013} used a large number of blocks to monitor the TE plant, constraining the variance captured by the local PCA models, but was unable to achieve a high detection rate in Fault 16. Based on this, we conclude that using fewer blocks and applying full PCA is a more effective approach.

When comparing CMAR-PCA with the centralized techniques, it is observed that this method can identify fault 5 as a back-to-control fault which KPCA and AE are unable to do. Furthermore, CMAR-PCA exhibits fault detection performance comparable to complex nonlinear methods such as OAE and BRNN
on back-to-control and uncontrollable faults. This suggests that employing a linear approach in a distributed manner yields results equivalent to centralized methods that account for process nonlinearity. Apparently, full PCA seems to outperform most other techniques. However, it is important to notice that for controllable faults, a high FDR can be detrimental to process monitoring. If operators conduct a thorough inspection of the plant, following an alarm on a controllable fault, they will likely find no evidence of anomalies in the process. Thus, emphasis is placed on techniques that achieve high detection rates for uncontrollable and back-to-control faults and low detection rates for controllable faults. In this study, by implementing full PCA in a distributed manner and performing the fault check through the Bayesian Inference Criterion (BIC), the FDR for controllable faults is close to the expected FAR for faults 3 and 9. For fault 15, the FDR is reduced compared to the centralized full PCA implementation.

\begin{figure*}[h] 
    \centering
    \includegraphics[width=0.8\textwidth]{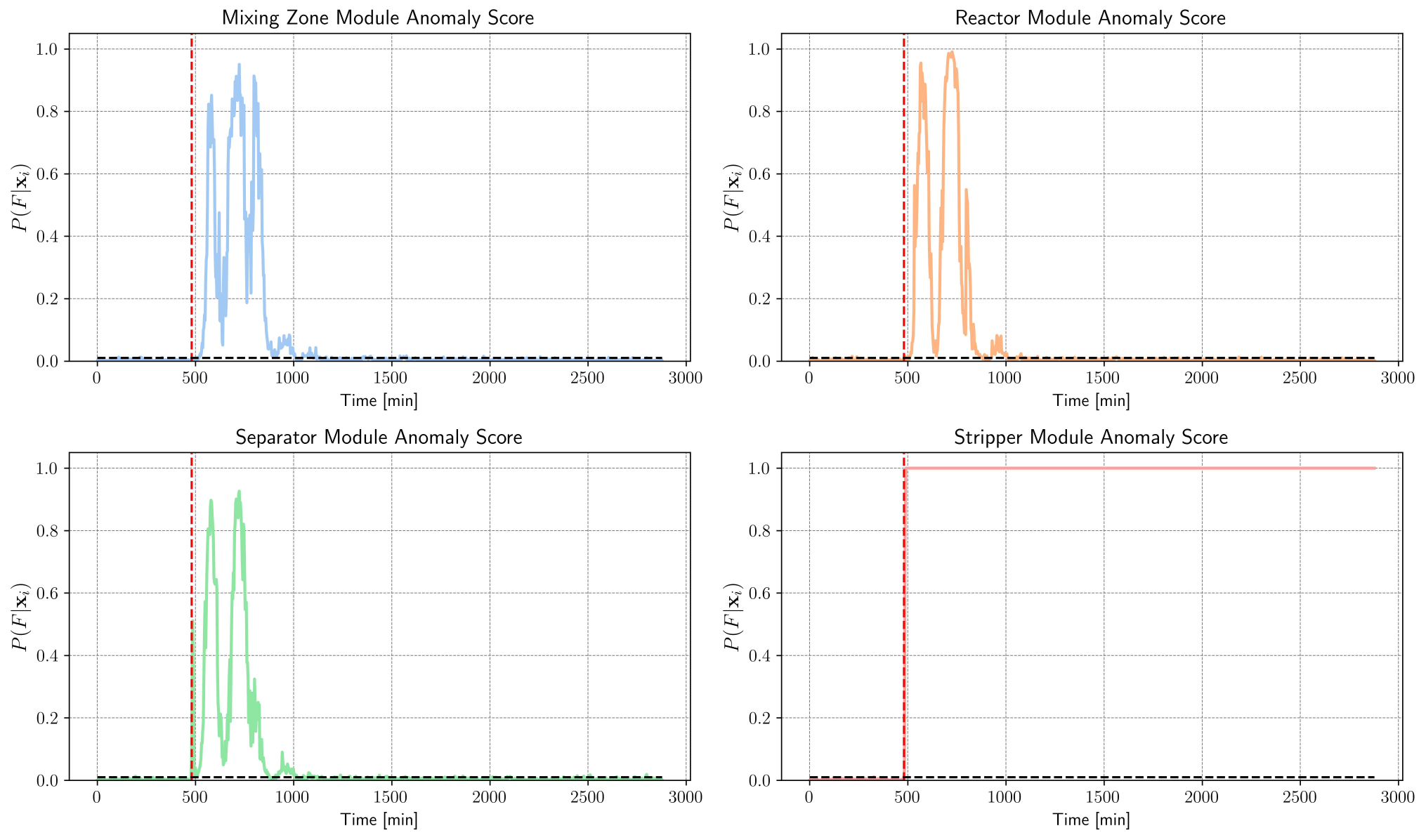} 
    \caption{Monitoring plots for Fault 5, organized into a visualization interface. The red dashed vertical line marks the fault onset, while the black horizontal dashed line represents the control limit of the anomaly score. \textbf{Top left:} Mixing Zone/Compressor monitoring block. \textbf{Top right:} Reactor monitoring block. \textbf{Bottom left:} Separator/Condenser monitoring block. \textbf{Bottom right:} Stripper monitoring block. The Y-axis shows the local anomaly score, representing the probability of fault given a new observation, and the X-axis represents time in minutes.}
    \label{fig:Mon5}
\end{figure*}

\subsection{Fault Diagnosis}
This section illustrates the application of the proposed framework for fault diagnosis on selected faults within the Tennessee Eastman Process (TEP) benchmark. Notice that in the monitoring plots the vertical red dashed line indicates the fault initiation, while the horizontal black dashed line represents the control limit.

\subsubsection{Fault 5: Step change in condenser cooling water inlet temperature at t = 480 min}
This fault is considered "back-to-control" \cite{Sun2020}, as the system non manipulated variables eventually return to the normal operating state. Fig. \ref{fig:Mon5} shows that all monitoring blocks obtained from the control aware MAR-based algorithm immediately flag the fault after it occurs. The blocks related to the mixing zone, reactor, and separator exhibit slight oscillations, which can be attributed to the control action attempting to restore the process to normal operation. Notably, the monitoring block corresponding to the stripper unit does not oscillate and maintains the anomaly flag, indicating that despite most parts of the process returning to control, the stripper continues to operate away from normal operation.

\begin{figure}[htbp] 
    \centering
    \includegraphics[width=0.45\textwidth]{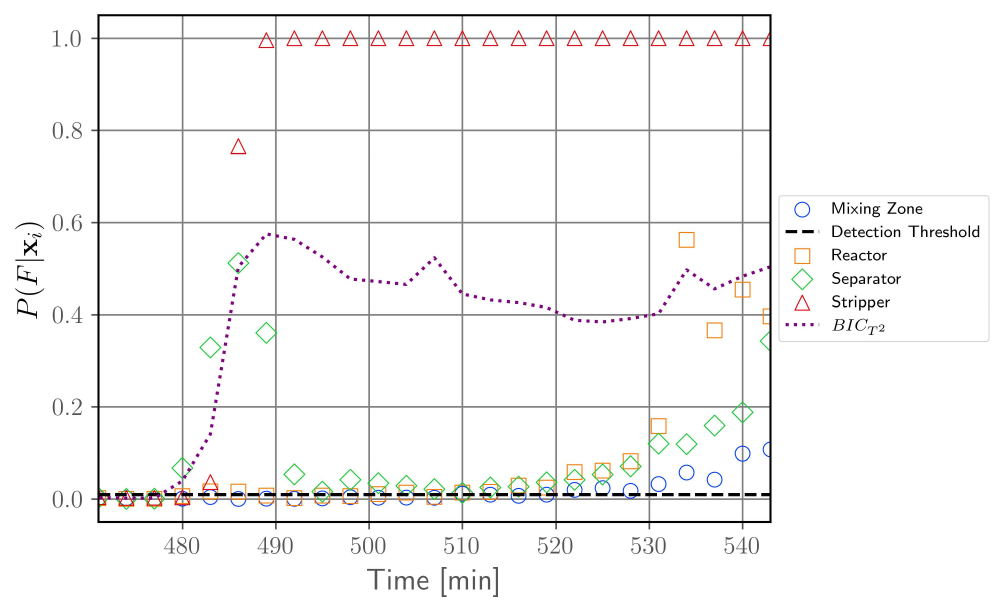} 
    \caption{Aggregate monitoring plot for Fault 5. Note that the separator block is the first to detect
the anomaly.}
    \label{fig:BIC5}
\end{figure}

\begin{figure}[htbp] 
    \centering
    \includegraphics[width=0.45\textwidth]{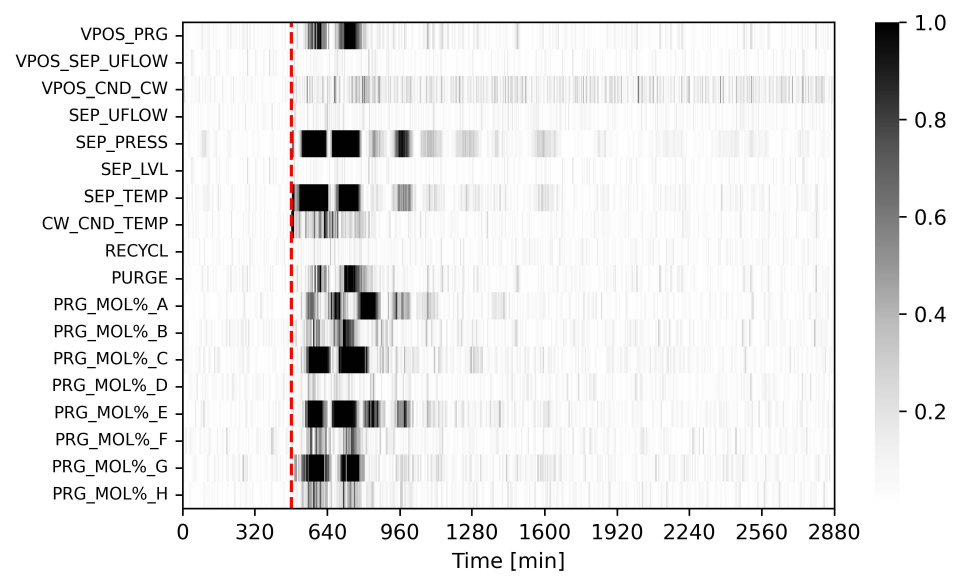} 
    \caption{2D Contributions plot for Fault 5 corresponding to the Separator monitoring block.}
    \label{fig:2D5}
\end{figure}

The alarm sequence reveals that the separator block triggered the alarm first, as illustrated in Fig. \ref{fig:BIC5}, which shows the initial fault propagation through all monitoring blocks. The contributions map corresponding to this block (Fig. \ref{fig:2D5}) indicates that the fault originated as a change of the condenser cooling water outlet temperature (XMEAS (22)). The root cause can be inferred as a change in the inlet water temperature, as there is no evidence of reactor problems affecting the product stream temperature or fouling in the condenser (since fouling cannot occur as a step change).

\begin{figure*}[h] 
    \centering
    \includegraphics[width=0.8\textwidth]{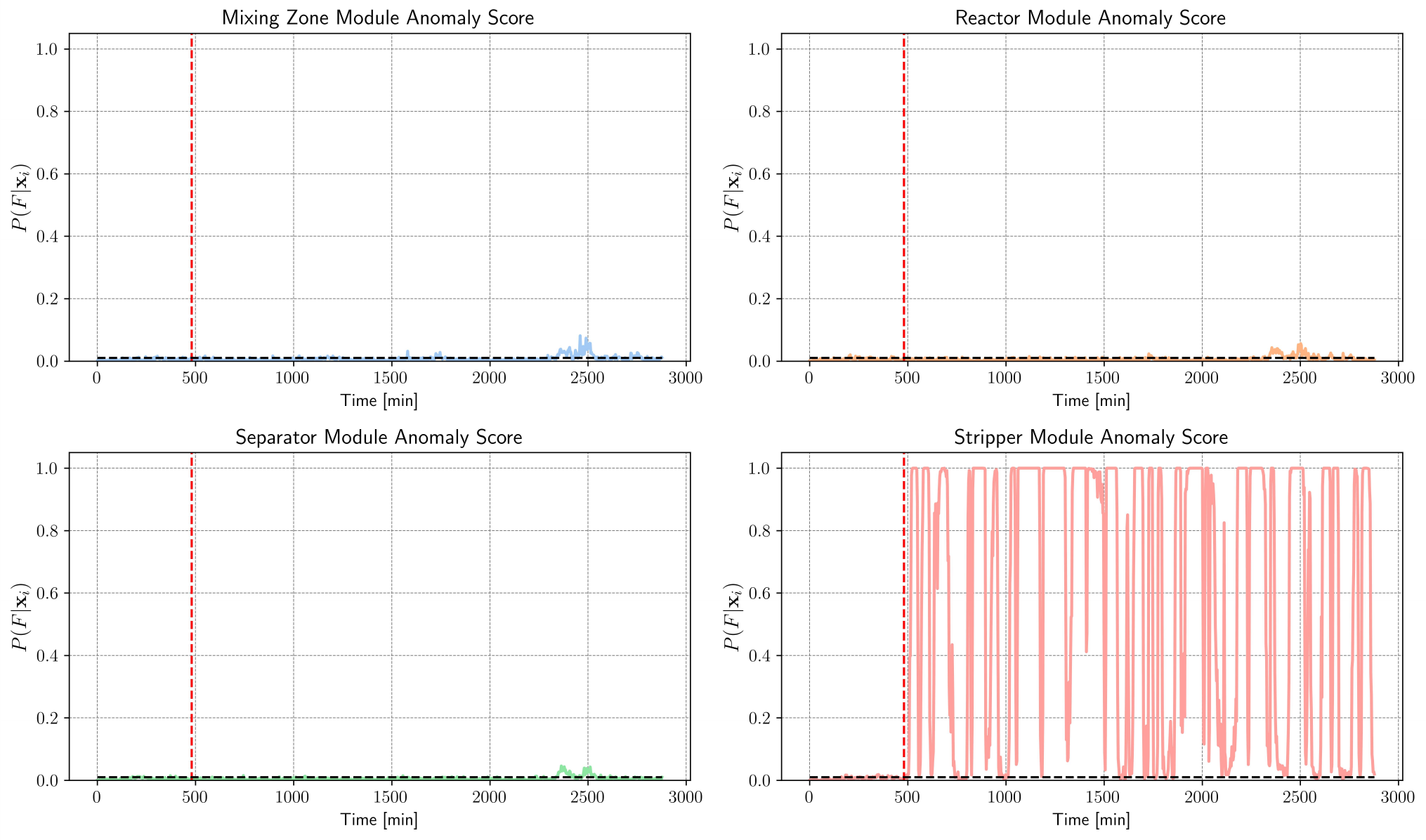} 
    \caption{Monitoring plots for Fault 16, organized into a visualization interface. The red dashed vertical line marks the fault onset, while the black horizontal dashed line represents the control limit of the anomaly score. \textbf{Top left:} Mixing Zone/Compressor monitoring block. \textbf{Top right:} Reactor monitoring block. \textbf{Bottom left:} Separator/Condenser monitoring block. \textbf{Bottom right:} Stripper monitoring block. The Y-axis shows the local anomaly score, representing the probability of fault given a new observation, and the X-axis represents time in minutes.}
    \label{fig:Mon16}
\end{figure*}

Furthermore, the plot shows that the valve position of the condenser cooling water (XMV (11)) remains above the normal operating region even after most variables return to control. This is logical since the fault impacts the amount of product condensed, which in turn affects the stripper flow in stream 11, leading to a change in the condenser cooling water flow. The inclusion of the manipulated variable of that control loop in the stripper block explains why the alarm remains flagged in the stripper block, showcasing the advantage of embedding control information into the monitoring blocks.

\subsection{Fault 16: Random deviations of heat transfer within the stripper}

\begin{figure}[htbp] 
    \centering
    \includegraphics[width=0.45\textwidth]{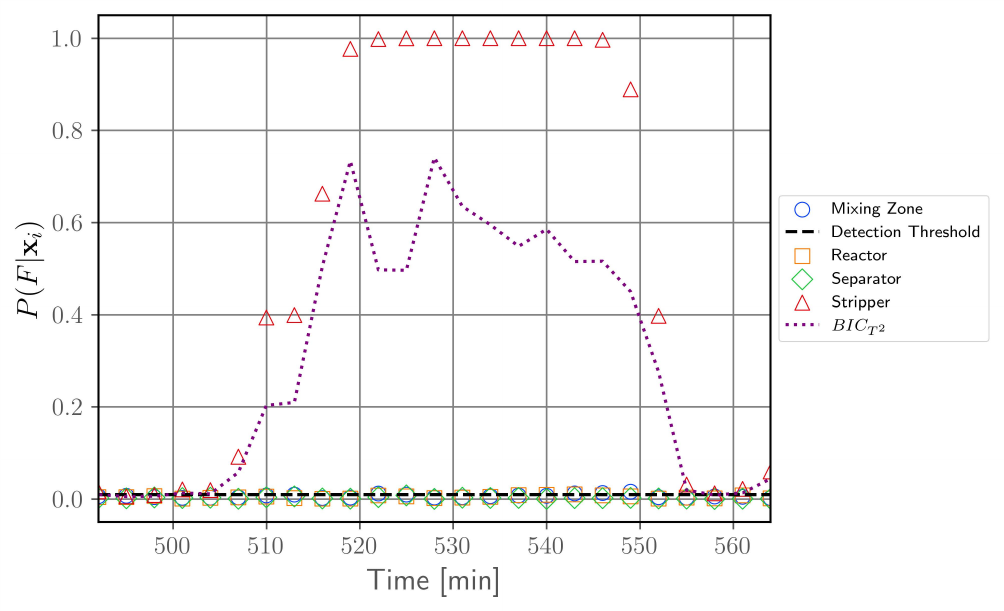} 
    \caption{Aggregate monitoring plot for Fault 16. Note that the stripper block is the first to
detect the anomaly.}
    \label{fig:BIC16}
\end{figure}

The monitoring plot (Fig. \ref{fig:Mon16}) indicates that the fault is initially contained within the stripper block, and the control system is attempting to compensate for it. If the control system remains triggered, it can be concluded that the fouling has not yet reached a critical phase. The activation sequence identifies that the fault originated in the stripper block (Fig. \ref{fig:BIC16}).

\begin{figure}[p] 
    \centering
    \includegraphics[width=0.45\textwidth]{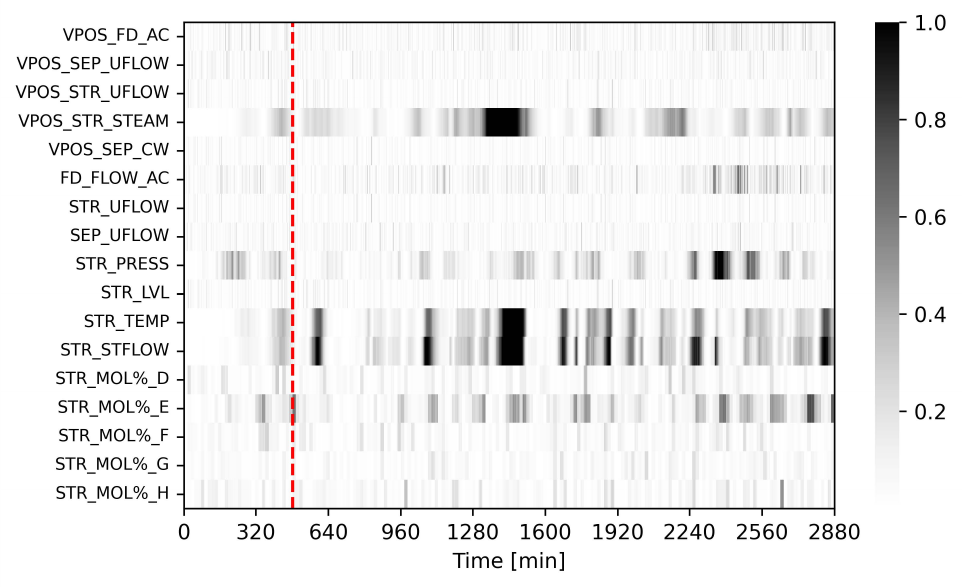} 
    \caption{2D Contributions plot for Fault 16 corresponding to the Stripper monitoring block.}
    \label{fig:2D16}
\end{figure}

Analyzing the contributions plot of this block (Fig. \ref{fig:2D16}), the noise does not allow for a clear distinction of the root cause of the fault. However, examining the patterns surrounding t = 480 min reveals that the valve position of the stripper steam (XMV (9)), stripper temperature (XMEAS (18)), stripper steam flow (XMEAS (19)), and concentration of E in stripper underflow (XMEAS(38)) enter the abnormal region in close succession. This indicates that something is triggering the control action in this cascade loop.

\begin{figure*}[htbp] 
    \centering
    \includegraphics[width=0.8\textwidth]{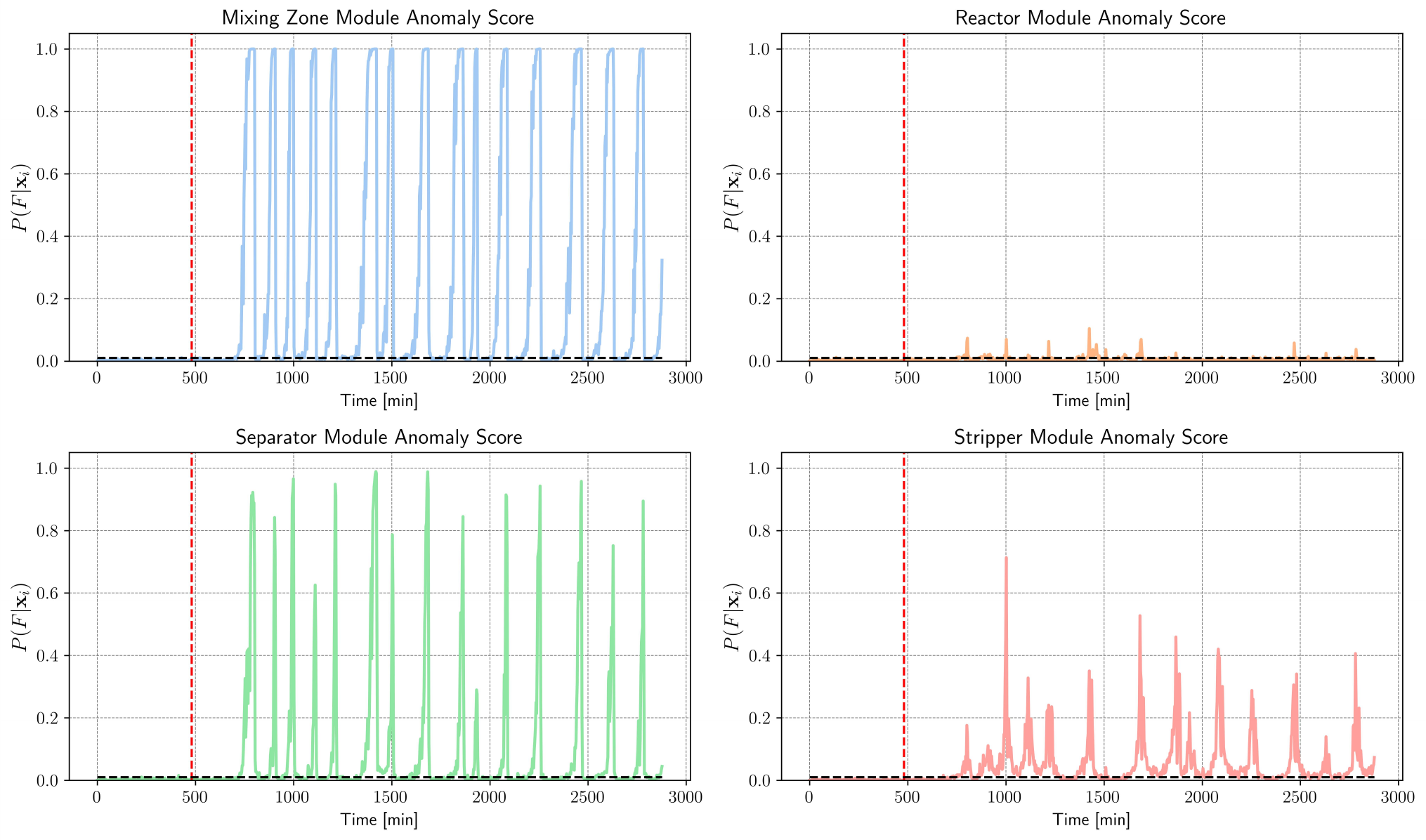} 
    \caption{Monitoring plots for Fault 20, organized into a visualization interface. The red dashed vertical line marks the fault onset, while the black horizontal dashed line represents the control limit of the anomaly score. \textbf{Top left:} Mixing Zone/Compressor monitoring block. \textbf{Top right:} Reactor monitoring block. \textbf{Bottom left:} Separator/Condenser monitoring block. \textbf{Bottom right:} Stripper monitoring block. The Y-axis shows the local anomaly score, representing the probability of fault given a new observation, and the X-axis represents time in minutes.}
    \label{fig:Mon20}
\end{figure*}

The only way this loop is triggered under these conditions is by an increase/decrease in the presence of component E in the product stream. The composition of component E in the effluent is heavily dependent on the temperature of the stripper. Given that variables in other blocks are close to their normal condition state, temperature changes in the inlets of the block can be ruled out. So, a control engineer might deduce that fouling is occurring and corroborate this by comparing the estimated heat transfer coefficient with the nominal that was calculated on the last maintenance of the process unit.

\subsection{Fault 20: Deviations of flow rate within stream 7}
\begin{figure}[htbp] 
    \centering
    \includegraphics[width=0.45\textwidth]{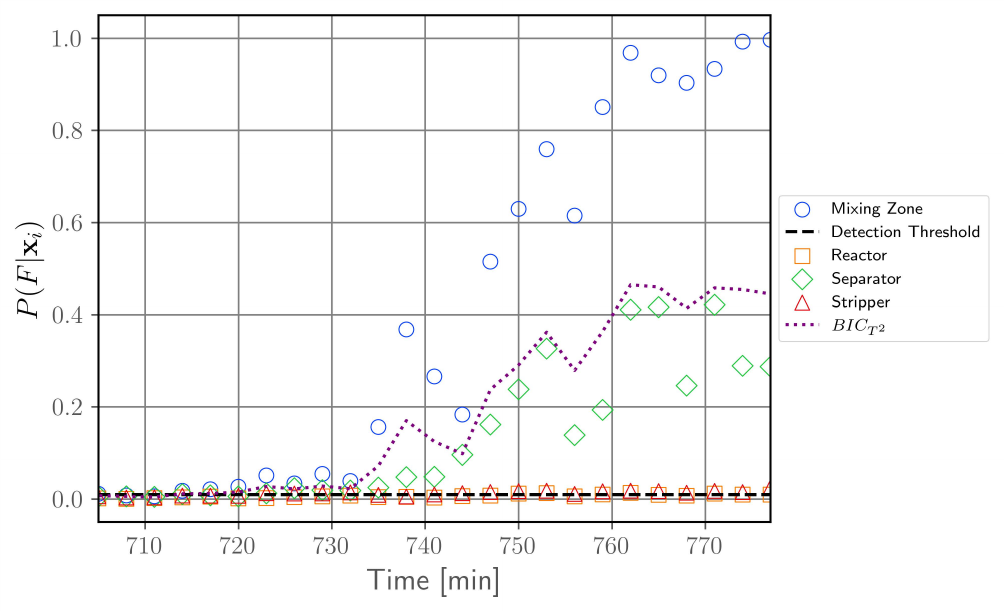} 
    \caption{Aggregate monitoring plot for Fault 20. Note that the mixing zone block is the first to detect the anomaly.}
    \label{fig:BIC20}
\end{figure}

Fault 20 is an uncontrollable fault labeled as unknown in the original TEP paper \cite{Downs1993} and was not unconcealed by Bathelt et al. \cite{Bathelt2015} in their revision of the code. By analyzing the FORTRAN
code, it was noticed that the fault consists in disturbances entering in Stream 7 (reactor’s effluent) flow rate. The flow of stream 7 is given by a flow equation that can be simplified as:

\begin{equation}
F_7 = C_v \sqrt{\Delta P},
\end{equation}

when fault 20 enters the system, the value of $C_v$ oscillates between 75\% and 100\% of its nominal value randomly. The monitoring plots (Fig. \ref{fig:Mon20}) showcase an oscillatory pattern in the mixing zone and separator blocks, which implies that the abnormal behavior is intermittent. The alarm sequence in this case indicates that the mixing zone monitoring block (Fig. \ref{fig:BIC20}) is the first to go off spec.

\begin{figure}[htbp] 
    \centering
    \includegraphics[width=0.45\textwidth]{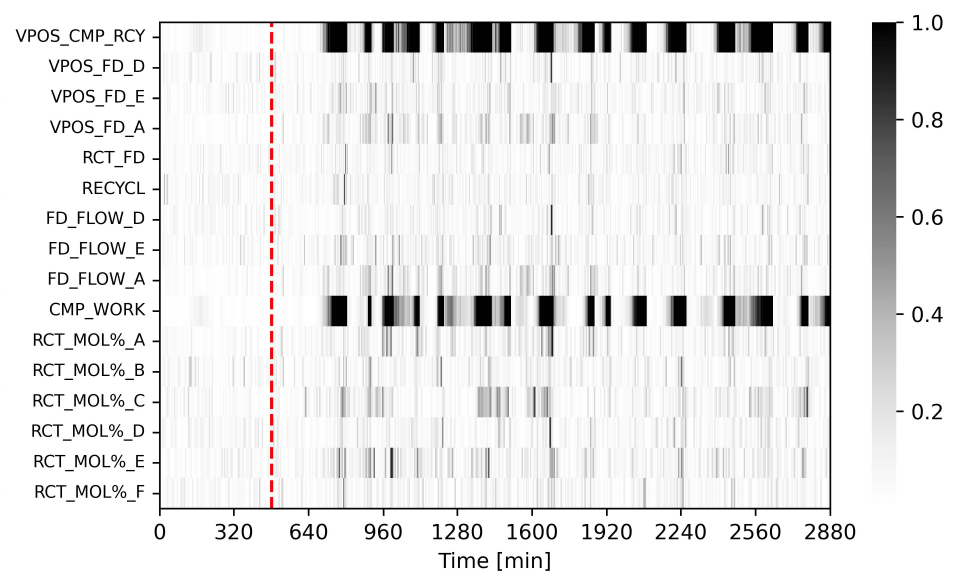} 
    \caption{2D Contribution plot for Fault 20 corresponding to the mixing zone monitoring block.}
    \label{fig:2D120}
\end{figure}

The contributions plot of the mixing zone monitoring block (Fig. \ref{fig:2D120}) shows that the valve position of the compressor recycle (XMV (5)) as well as the compressor work (XMEAS (20)) are in the abnormal operating region. A decrease in the flow of Stream 7 causes changes in the recycle flow which ultimately triggers the control action that moves the position of the compressor recycle valve in order to keep the recycle flow at the setpoint. This is perfectly illustrated in Fig. \ref{fig:2D120} as the other variables’ contribution to the anomaly score is relatively low most of the time.

\begin{figure}[htbp] 
    \centering
    \includegraphics[width=0.45\textwidth]{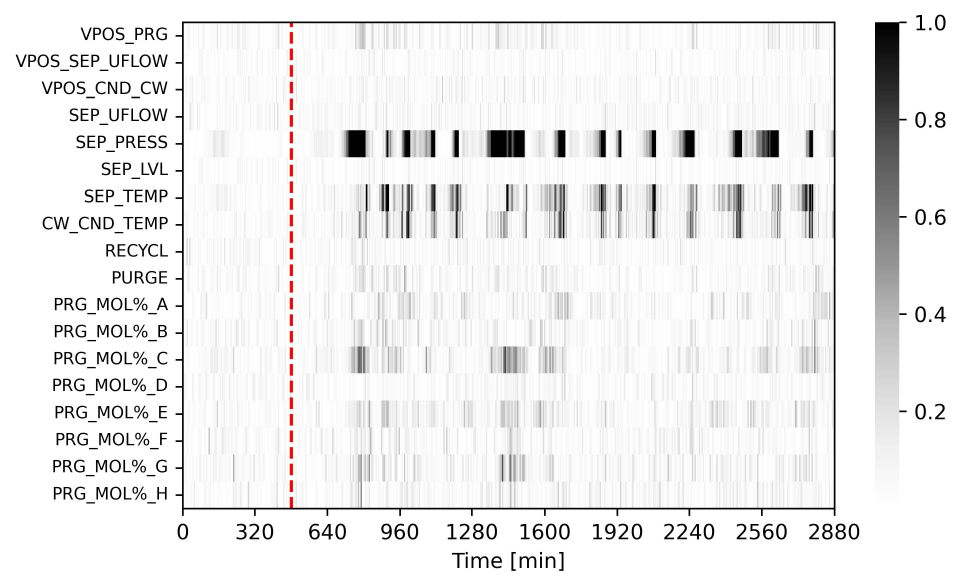} 
    \caption{2D Contribution plot for Fault 20 corresponding to the separator monitoring block.}
    \label{fig:2D220}
\end{figure}

In this case it is also worth looking at the separator block contributions plot (Fig. \ref{fig:2D220}) since this block also exceeds the control limit in close succession of the mixing zone block. The contribution plot reveals that the first variable to deviate from the normal operating conditions is the separator’s pressure (XMEAS (13)), as the three control valves related to that block are still on their normal operation positions, it is possible to conclude that the pressure change is due to a change in the vapor being produced in this separator this information along with the one obtained from the mixing zone contributions plot allows a control engineer to infer that the cause of this anomalies is a decrease in the feed flowrate of the separator.

\subsection{Fault propagation analysis}
\begin{figure*}[t] 
    \centering
    \includegraphics[width=0.8\textwidth]{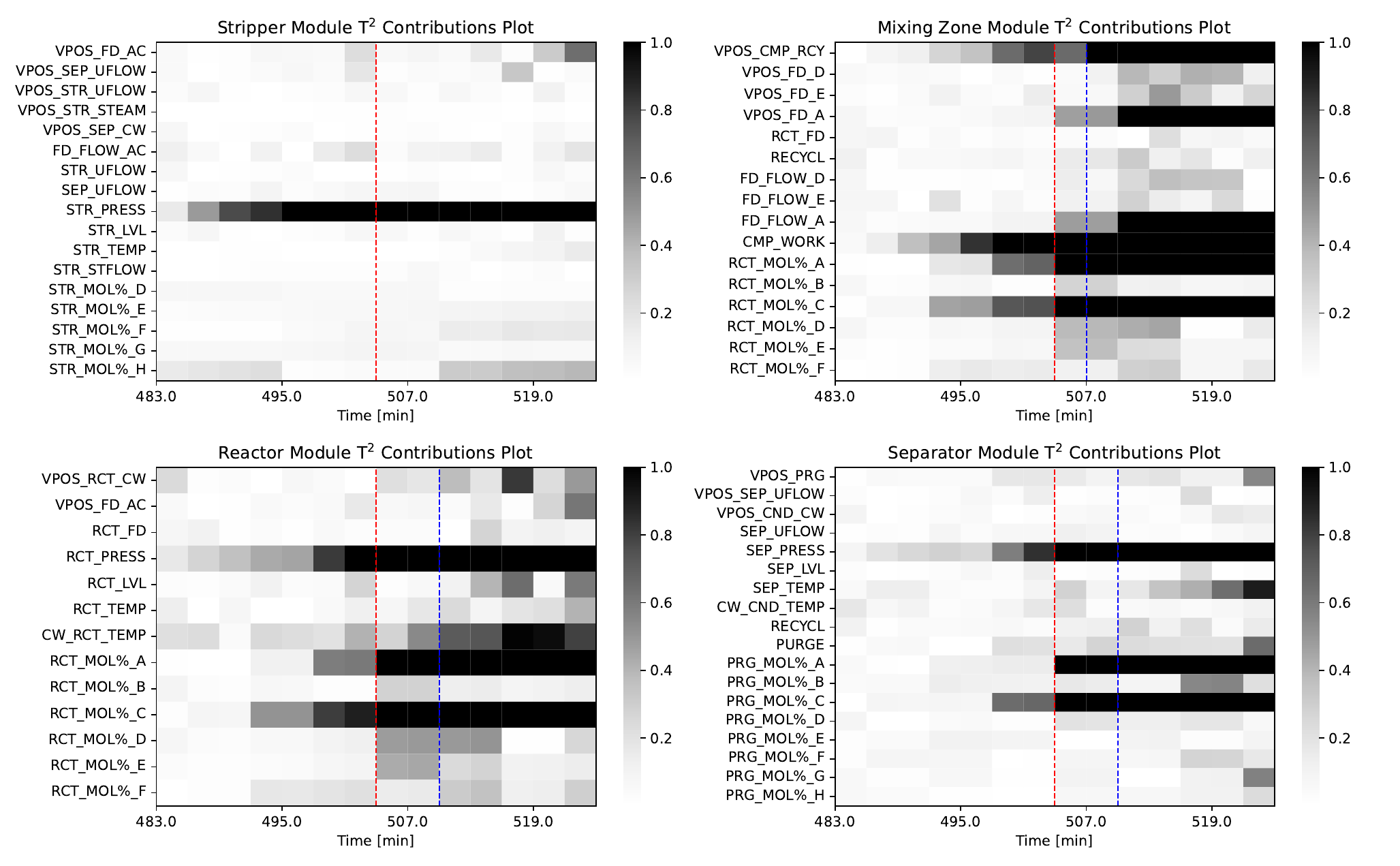} 
    \caption{2D contribution plot sequence for Fault 1. The vertical dashed line indicates the fault onset, while the blue dashed line marks the moment the fault was detected by the local anomaly score in each block. Scaling and clipping of the contributions enhance the analysis of fault propagation within the plant. }
    \label{fig:prop}
\end{figure*}

By generating the 2D contributions plot using equation (18) for each monitoring block, it is possible to compare them within the same time horizon where the fault has been detected, providing insight into the short-term effects of the fault. Consequently, in certain scenarios, a qualitative fault propagation analysis can be conducted through visual inspection of the contribution plots arranged according to the block’s alarm activation sequence (see Fig. \ref{fig:prop}). In this section, this propagation analysis is illustrated for Fault 1, which corresponds to a step change in the A/C-ratio of stream 4 at t = 480 min.

\begin{figure}[htbp] 
    \centering
    \includegraphics[width=0.45\textwidth]{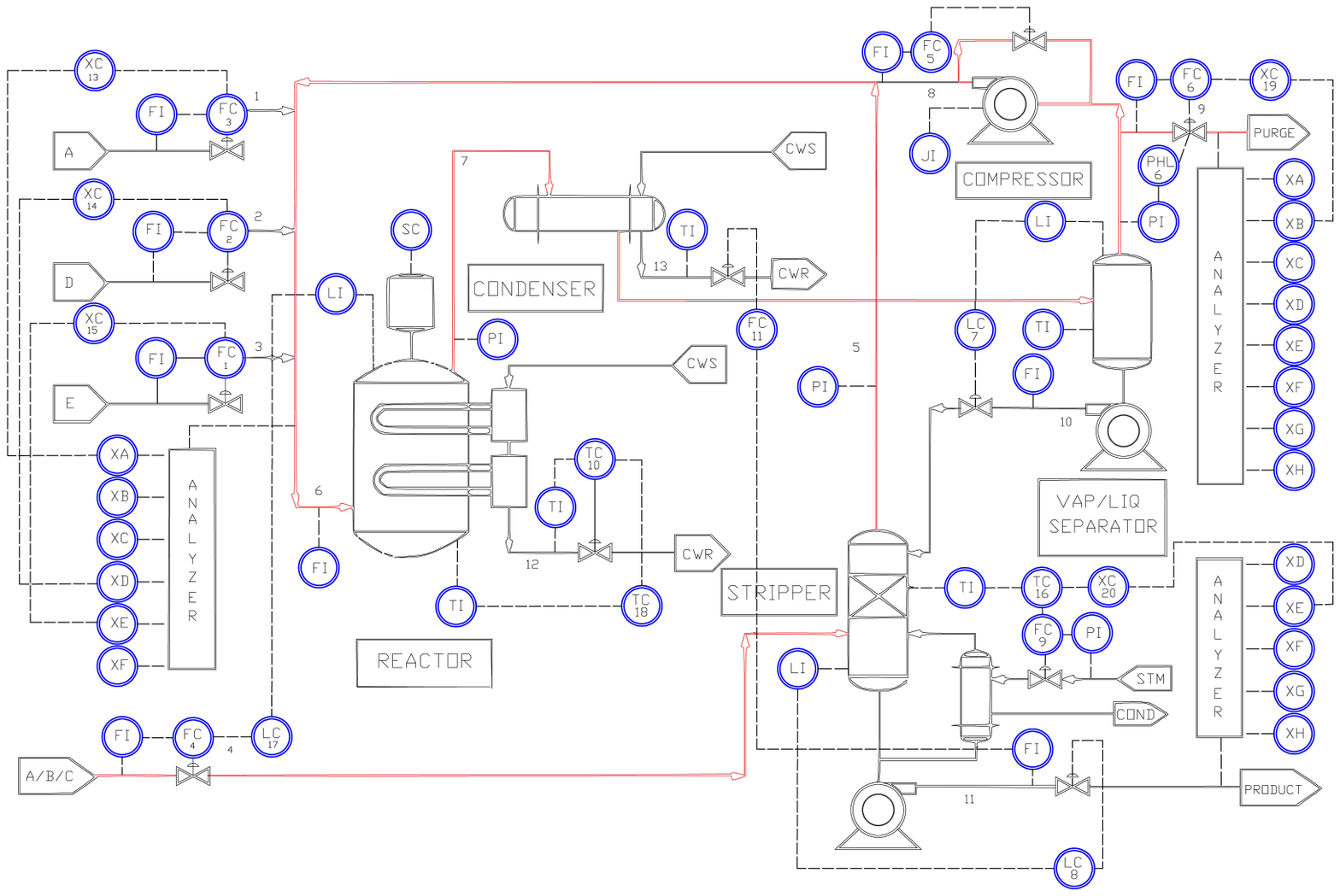} 
    \caption{Propagation path of Fault 1. The red path highlights the early stages of fault propagation through the plant.}
    \label{fig:prop2}
\end{figure}

From Fig. \ref{fig:prop} it can be seen that the fault is first detected in the stripper, then moves to the mixing zone, affects the reactor, and finally impacts the separator. Note that the red dashed line indicates the time at which the fault was first detected, and the blue line indicates the time at which the fault was detected in each respective monitoring block. Initially, the pressure in the stripper deviates from the normal operating state, and this change propagates to the mixing block. This propagation occurs through stream 5, the main source of component C to the reactor. The mixing zone contributions plot indicates that the control action has been triggered to keep component A concentration in the reactor feed at its setpoint, and it also shows that component C concentration has deviated from normal operation. The reactor contributions plot reveals that the changes in the reactor feed composition have affected the reactor’s pressure. As the reactor pressure changes, the reactor effluent flow is also altered. Initially, the condenser will attempt to condense the effluent amount corresponding to normal operation, leading to changes in the separator’s pressure. The separator block contributions plot shows that the concentrations of components A and C in the process have ultimately changed. As the amounts of reactants A and C vary, the compressor’s recycle valve control action attempts to maintain the recycle flow at the setpoint, which is clearly shown in the mixing zone contributions plot. From this analysis, it is inferred that Fault 1 propagates in its early stages as depicted in Figure \ref{fig:prop2}.

\section{Conclusions} \label{sec:conc}

This work introduces a novel knowledge-based process decomposition algorithm for decentralized process monitoring that is based on PFDs and the control loop structure. The algorithm identifies decomposition blocks of interacting process variables, resulting in blocks that are explainable and more suited to the process structure than the blocks identified by prior decomposition methods \cite{Ge2013,Chen2019,Peng2020}. The proposed algorithm does not require information about the correlation or mutual information among process variables to create monitoring blocks for effective fault detection. Community merging based on measurement allocation has been shown to lead to a superior definition of monitoring blocks when compared to the standard knowledge-based process decomposition algorithms that do not typically consider plant sensor placement as a criterion for merging communities.

The goal of this work is to introduce an FDI methodology that is on par with the most sophisticated recent advances in FDI while employing algorithms that are within the reach of an average control engineer, without requiring specialized training. The methodology consists in performing the process decomposition based on the PFD, merging blocks with few process variables based on measurement allocation, refining the blocks using control loop information and constructing an f-PCA model in each monitoring block. A Bayesian aggregate fault detection index which combines fault detection indices from individual blocks into an overall plant level assessment index along with a fault tracking visualization interface for variables within each block, enables visual assessment of faults and their progress through time.

The results demonstrate that by defining monitoring blocks based on the plant sensor placement and the control loops, modular full PCA can detect faults as effectively as more complex nonlinear methods or centralized full PCA. Additionally, the distributed process monitoring approach simplifies the fault identification step for certain faults. For many examples of faults in TEP, propagation of fault detection through individual process monitoring blocks clearly shows the block where the fault originates. The modified contribution plots proposed in this work for individual blocks show clearly the sequence of measured and controlled variable changes after the fault occurs. Since the number of variables associated with each monitoring block is relatively small (around 15), it is straightforward to identify the fault or to identify alternative fault candidates which then need to be verified further in the plant. Relative simplicity of the method, excellent results in fault detection, and straightforward identification through analysis of fault propagation and control actions, make the method very suitable for plant implementation.

\section{Acknowledgments}
This work was supported by the McMaster Advanced Control Consortium (MACC).
\section{CRediT authorship contribution statement}
Enrique Luna Villagómez: Conceptualization, Methodology, Software, Writing – original
draft. Vladimir Mahalec: Supervision, Conceptualization, Methodology, Writing – review \&
editing.






\bibliographystyle{elsarticle-num-names} 
\bibliography{cas-refs}






\end{document}